\documentclass[%
,secnumarabic%
,tightenlines%
,amssymb, amsmath,nobibnotes, aps,nofootinbib,showpacs,showkeys]{revtex4}
\usepackage{epsfig}%
\usepackage{graphicx}%
\expandafter\ifx\csname package@font\endcsname\relax\else
 \expandafter\expandafter
 \expandafter\usepackage
 \expandafter\expandafter
 \expandafter{\csname package@font\endcsname}%
\fi

\begin{document}
  \title{Propagation of quantum particles in Brans--Dicke spacetime. The case of  Gamma Ray Bursts}
\author{S. Capozziello$^{a,b,c}$, G. Lambiase$^{d,e}$}
\affiliation{
\mbox{$^a$Dipartimento di Fisica, Universit\'a di Napoli "Federico II", I-80126 - Napoli, Italy.}\\
\mbox{$^b$INFN Sez. di Napoli, Compl. Univ. di Monte S. Angelo, Edificio G, I-80126 - Napoli, Italy,}\\
\mbox{$^c$Gran Sasso Science Institute (INFN), Viale F. Crispi, 7, I-67100, L'Aquila, Italy,}\\
\mbox{$^d$Dipartimento di Fisica "E.R. Caianiello" Universit\'a di Salerno, I-84081 Lancusi (Sa), Italy,}\\
\mbox{$^e$INFN - Gruppo Collegato di Salerno, Italy.}
}

\date{\today}

\renewcommand{\theequation}{\thesection.\arabic{equation}}
\begin{abstract}
The propagation of  boson particles in a gravitational field
described by the Brans-Dicke theory of gravity is analyzed. We
derive the wave function of the scalar particles, and the effective
potential experienced by the quantum particles considering the role of the varying gravitational coupling. Besides, we
calculate the probability to find the scalar particles near the
region where a naked singularity is present. The extremely high  energy radiated in such a situation could account for the huge emitted power observed 
in Gamma Ray Bursts.

\end{abstract}

\maketitle

\section{Introduction}
\setcounter{equation}{0}

The  Brans--Dicke (BD) Theory  \cite{brans} provides an
extension of Einstein's General Relativity  where  Mach's principle and
Dirac's large number hypothesis  \cite{weinberg} are properly accommodated by means of a
nonminimal coupling between the geometry and a scalar field $\phi$.
Such a theory is the prototype of any scalar-tensor gravity and, in general, of any Extended Theory  of Gravity \cite{rept,odintsov}.

Some concomitant circumstances have  renewed interest in the BD
theory and in its scalar-tensor extensions. These are: $i)$ The unavoidable presence of scalar fields
in superstring theory and in other unified scheme \cite{green};
$ii)$ the fact that scalar-tensor gravity can be recovered from the
Kaluza-Klein theory once extra dimensions are compactified
\cite{cho}; $iii)$ the BD theory plays an important role in
extended and hyperextended inflationary scenarios
\cite{la,kolb,steinardt,liddle}; $iv)$ some astronomical  observations
\cite{pioneer,nieto} seem to ask  for a  space-time variation of the Newton
gravitational constant on astrophysical and cosmological  scales. Similar variations are predicted by the BD
theory which has also consequences for neutrino oscillations
\cite{gaetano}, the gravity induced Cerenkov effect
\cite{cerenkov}, and Sagnac effect \cite{nandi}.

A very interesting feature of the BD theory is that the scalar
field generates a singularity at the horizon, that prevents
collapse and formation of a black hole in a stationary,
spherically symmetric geometry \cite{janis,agnese}. This unusual
position of the singularity seems contrary to Penrose's cosmic
censorship hypothesis \cite{penrose} whereby singularities in
General Relativity are always {\it clothed} by event horizons and
cannot therefore be visible from  outside.  This hypothesis
is of great importance in the study of the global structure of
 space-time. In its absence, the future evolution of
space-times regions containing naked singularities cannot be
determined since new information could emerge from there in
completely arbitrary ways. The validity of the cosmic censorship
hypothesis is still an open issue in General Relativity. A few
example of naked singularity formation can be given. They refer
to gravitational collapse of dust and imperfect fluids
\cite{singh}. Typically, for a given equation of state, both
black holes and naked singularities can occur, depending on the
choice of the initial conditions. However, the most striking
evidence of a violation of Penrose's hypothesis comes from the
study of collapse in the presence of massless scalar fields
\cite{janis,agnese,choptuik}).

The purpose of this paper is to analyze the effect of
singularities at the event horizon on the propagation of
particles described by the Klein-Gordon equation. The background
is represented by a static, spherically symmetric solution of the 
BD theory. We derive the effective potential experienced by a
particle in this  background and show that it gives rise
to an infinitely attractive force at the singularity so that the
particles cannot approach the origin. This leads to the
conclusion that the presence of $\phi$ prevents the formation of
black holes, as argued in \cite{agnese}. We then derive the
electromagnetic power emitted by the sclar particle propagating
in the BD background, and show that, for particular values of the
BD parameters, it can be of the order of $10^{54}$erg/s and
therefore comparable with the power emitted by gamma ray bursts (GRBs).
It is important to stress the role of   nonminimal  gravitational coupling is determining such a dynamics.

The paper is organized as follows. In Section II we summarize
the main features of BD theory of gravity and the solution of the
field equation corresponding to a static source. Then, we
investigate the structure of the singularity. In Section III we
write down the wave equation of a scalar particle, and derive
the effective potential. In Section IV, the probability to find
the scalar particle near $r=B$ is calculated. In Section V, these
results are used for studying the GRBs physics. Conclusions are
drawn in Section VI.

\section{Static Spherically Symmetric Solutions in Brans--Dicke Theory}
\setcounter{equation}{0}

In BD theory, the effective action describing the interaction of a scalar
field $\phi$ nonminimally coupled with the geometry and the
ordinary matter is given by \cite{brans}
\begin{equation}\label{1.1}
{\cal A}=\int d^4x \sqrt{-g}\left[\phi {\cal R}[g]-\omega
\frac{\partial_{\mu}\phi\partial^{\mu}\phi}{\phi}+\frac{16\pi}{c^4}
{\cal L}_{m}\right],
\end{equation}
where ${\cal R}[g]$ is the scalar curvature function of the metric $g_{\mu\nu}$, and  ${\cal L}_{m}$ is the ordinary
matter contribution. The constant
$\omega$ is determined by observations and its value can be
constrained by classical tests of General Relativity. The
consequences of BD action (\ref{1.1}) have been analyzed for the
light deflection, the relativistic perihelion rotation of Mercury,
and the time delay experiments, resulting in reasonable agreement
with all available observations thus far provided ($\omega \geq
500$)\cite{will}. The most recent bound $\omega>3000$ has been
derived by Very-Long Baseline Radio Interferometry (VLBI)
experiment \cite{will-rev}. On the other hand, bounds on the
anisotropy of the microwave background radiation give the upper
limit $\omega\leq 30$\footnote{This dichotomy in the bounds of
$\omega$ of the
original BD scheme has led  to take into account self-interaction
scalar field potential (see e.g. \cite{salva,lobo}.)}\cite{la}.
Nevertheless, as shown in Refs. \cite{xue,xue2}, the consequences of
the BD theory in the radiation-matter equality, seem to indicate
that at the moment there is no observable effects appearing from
CMBR. Einstein's theory is fully recovered for $\omega\to\infty$. In this
limit, the BD theory becomes indistinguishable from General
Relativity in all its predictions.
The variation of the action (\ref{1.1}) with respect to the
metric tensor $g_{\mu\nu}$ and the scalar field $\phi$ yields 
the field equations \cite{brans}
\begin{equation}\label{2.1}
  {\cal R}_{\mu\nu}-\frac{1}{2}\,g_{\mu\nu}\, {\cal R} = \frac{8\pi }{c^4\phi}\, T_{\mu\nu}+
  \frac{\omega}{\phi^2}\left(\phi_{,\mu}\phi_{,\nu}-\frac{1}{2} g_{\mu\nu}
  \phi_{,\alpha}\phi^{,\alpha}\right)
   +\frac{1}{\phi}\,(\phi_{,\mu ;\nu}-g_{\mu\nu}\Box \phi)
\end{equation}
and
\begin{equation}\label{2.2}
  \frac{2\omega}{\phi}\,\Box \phi-\frac{\omega}{\phi^2} \phi_{,\mu}\phi^{,\mu}
  +{\cal R}=0\,.
\end{equation}
Here $\Box$ is the  d'Alembert operator in curved
spacetime.

$T_{\mu\nu}$ is the energy-momentum tensor of matter. The line
element describing a static and isotropic geometry is
\begin{equation}\label{2.3}
  ds^2=-e^{\nu}dt^2+e^{\mu}[dr^2+r^2(d\theta^2+\sin^2\theta d\varphi^2)]\,,
\end{equation}
where the functions $\nu$ and $\mu$ depend on the radial
coordinate $r$. For $r>B$, the general solution in the vacuum is
given by
\begin{eqnarray}
 e^{\nu} & = & e^{2\alpha_0}\left[\frac{1-B/r}{1+B/r}\right]^{2/\lambda}\,,\label{2.4} \\
 e^{\mu} & = & e^{2\beta_0}\left(1+\frac{B}{r}\right)^4
 \left[\frac{1-B/r}{1+B/r}\right]^{2(\lambda -C-1)/\lambda}\,, \label{2.5} \\
 \phi & = & \phi_0\left[\frac{1-B/r}{1+B/r}\right]^{-C/\lambda} \,, \label{2.6}
\end{eqnarray}
where $\alpha_0$, $\beta_0$, $\phi_0$, $B$, and $C$ are arbitrary
constants. $B$ is related to the mass of the source. Here
$\omega>3/2$ and
\begin{equation}\label{lambda}
  \lambda= \left[(C+1)^2-C\left(1-\frac{\omega C}{2}\right)\right]^{1/2}\,.
\end{equation}
In the weak field approximation, the constraints on BD parameters
are \cite{brans}
\begin{equation}\label{2.7}
  \lambda =\sqrt{\frac{2\omega +3}{2(\omega +2)}}\,, \quad C\cong -\frac{1}{2+\omega}\,,
  \quad \alpha_0=0=\beta_0\,,
\end{equation}
 \begin{equation}
 \phi_0=\frac{4+2\omega}{G_N(3+2\omega)}\,, \quad B=\frac{M}{2c^2\phi_0}\,
 \sqrt{\frac{2\omega+4}{2\omega+3}}\,.
\end{equation}
Clearly the mass and the gravitational coupling depend on the value of the scalar field.
Notice that the metric (\ref{2.3}) has a naked singularity at the
point $r=B$ (the singularity is naked because it is not covered by
the event horizon). In fact, the  Kretschmann scalar  invariant
 \[
 {\cal R}_{\alpha\beta\rho\delta}{\cal R}^{\alpha\beta\rho\delta}=
 \frac{16B^2r^6}{(r^2-B^2)^8\lambda^4}\left(\frac{r-B}{r+B}\right)^{4(1+C)/\lambda}
 \left\{3(B^4+r^4)(2+2C+C^2)[2+2C+C^2(2+\omega)] \right.
 \]
 \begin{equation}
 \left. -8B\lambda
 r(B^2+r^2)[6+9C+C^3(3+\omega)+C^2(7+\omega)]+2B^2r^2[60C+6(5+\lambda^4)
 \right.
 \end{equation}
 \[
 \left. +8C^2(8+\omega)+C^4(12+5\omega)+
 2C^3(19+5\omega)]\right\}
 \,,
 \]
as well as the scalar curvature
 \begin{equation}
 {\cal R}=\frac{4B^2}{(r^2-B^2)^4\lambda^2}\left(\frac{r-B}{r+B}\right)^{2(1+C)/\lambda}
 \omega C^2\,,
 \end{equation}
diverge at $r=B$, since
 \begin{equation}\label{l>c+1}
 \lambda>\frac{C+1}{2}\,,
 \end{equation}
as immediately follows from Eq.(\ref{lambda}).
To investigate the structure of the singularity, we examine the
properties of the equipotential surfaces $g_{00}=$constant,
$t=$constant, and the closed curves on the surfaces as $r\to B$
\cite{agnese}. We find that the area of such equipotential
surfaces is
 \begin{equation}
 A=\int_0^{2\pi}\int_0^\pi\sqrt{g_{\theta\theta}g_{\varphi\varphi}}\,d\theta d\varphi
 =4\pi r^2\left(1+\frac{B}{r}\right)^4
 \left[\frac{r-B}{r+B}\right]^{2(\lambda -C-1)/\lambda}\,.
 \end{equation}
The proper lengths for a closed azimutal curve $\theta=\pi/2$ is
 \begin{equation}
 L_\varphi=\int_0^{2\pi}\sqrt{-g_{\varphi\varphi}}\,d\varphi=2\pi r
 \left(1+\frac{B}{r}\right)^2
 \left[\frac{r-B}{r+B}\right]^{(\lambda -C-1)/\lambda}\,,
 \end{equation}
whereas for polar curve $\varphi=$constant, it is
\begin{equation}
 L_\theta=2\int_0^{\pi}\sqrt{-g_{\theta\theta}}\,d\theta=2\pi r
 \left(1+\frac{B}{r}\right)^2
 \left[\frac{r-B}{r+B}\right]^{(\lambda -C-1)/\lambda}\,.
 \end{equation}
It is evident that the singularity at $r=B$ has the topology of a
point when $\lambda>C+1$, provided $C>2/\omega$, for positive $C$,
or $C<0$. As a consequence, the event horizon shrinks  to a
point  and the collapse through the singularity becomes
impossible, preventing the formation of black holes \cite{agnese}.
Notice also that for $0<C<2/\omega$, the above quantities $A$,
$L_\varphi$, and $L_\theta$ are meaningless being divergent.

In the next Section we will investigate the propagation of a
scalar particle in the BD solutions (\ref{2.4})--(\ref{2.6}),
corresponding to the outside spherical surface of radius $r=B$.

\section{Klein-Gordon Equation and the Effective Potential}
\setcounter{equation}{0}

In a curved space--time, the wave equation for a scalar particle
of rest mass $m$ and minimally coupled to gravity is
\begin{equation}\label{weq}
(\Box + m^2)\psi(x)=0\,.
\end{equation}
 If the components of
the metric tensor are given by (\ref{2.3}), then Eq. (\ref{weq})
assumes the form
 \begin{equation}\label{weq1}
 \left\{\frac{\partial^2}{\partial t^2}
 -\frac{e^{(\nu-3\mu)/2}} {r^2}\frac{\partial} {\partial r}\left(
 e^{(\nu+\mu)/2} r^2\frac{\partial}{\partial r}\right) 
 - \frac{e^{\nu}}{r^2}\left[\frac{1}{\sin\theta}\frac{\partial}
 {\partial\theta}\left(\sin\theta\frac{\partial}{\partial\theta}\right)+
 \frac{1}{\sin^2\theta}\frac{\partial^2}{\partial\varphi^2}\right]
 - m^2e^{\nu}\right\}\psi(t,r,\theta,\varphi)=0\,{.}
 \end{equation}
By using the method of  separable variables, the wave function
can be written as
\begin{equation}\label{sep}
\psi(t,r,\theta,\varphi) = T(t)R(r)\Theta(\theta,\varphi)
\end{equation}
and Eq. (\ref{weq1}) splits into the following three equations
\begin{equation}\label{eqT}
\frac{\partial^2 T}{\partial t^2} + \Omega^2 T = 0\,{,}
\end{equation}
\begin{equation}\label{eqO}
\frac{1}{\Theta}\left[\frac{1}{\sin\theta}\frac{\partial}{\partial\theta}
\left(\sin\theta\frac{\partial}{\partial\theta}\right)+\frac{1}{\sin^2\theta}
\frac{\partial^2}{\partial\varphi^2}\right]\Theta = -l(l+1) \,{,}
\end{equation}
\begin{equation}\label{isoR}
e^{\nu-\mu}R^{\prime\prime}+e^{\nu-\mu}\left( \frac{2}{r}+
\frac{\nu^{\prime}}{2}+\frac{\mu^{\prime}}{2}\right)R^{\prime}+
 \left[\Omega^2-e^{\nu}
 \left(e^{-\mu}\frac{l(l+1)}{r^2}+m^2\right)\right]R=0\,{.}
\end{equation}
where $\Omega^2$ is a separation constant corresponding to the
frequency of the wave, $l$ is the  orbital angular quantum  momentum of the scalar particle and the prime indicates derivatives
with respect to $r$.

The solution of Eq. (\ref{eqO}) is
\begin{equation}\label{harm}
\Theta_{lp}(\theta,\varphi)=Y^p_l(\cos\theta)e^{ip\varphi}\,,
\end{equation}
where $Y^p_l(\cos\theta)$ are the usual spherical harmonics, and
$p$, with $\mid p \mid \leq l$, is the magnetic quantum number.
The general solution of Eq. (\ref{eqT}) is
\begin{equation}\label{temp}
T(t)=C_1e^{-i\Omega t}+C_2e^{i\Omega t}\,{,}
\end{equation}
where $C_1$ and $C_2$ are arbitrary constants. From Eqs.
(\ref{sep}), (\ref{harm}) and (\ref{temp}), it follows that the
eigenfunctions of the scalar wave equation, Eq. (\ref{weq1}), can
be recast in the form
\begin{equation}\label{eig}
\psi(t,r,\theta,\varphi)=R(r)Y_l^p(\cos\theta)e^{i(p\varphi\pm\Omega
t)}\,{,}
\end{equation}
where $R(r)$ is the solution of the radial wave equation
(\ref{isoR}).

In order to derive the effective potential from (\ref{isoR}), we
introduce the variable $r^*=r^*(r)$ such that
\begin{equation}\label{isonew}
e^{\nu-\mu}\left(\frac{dr^*}{dr}\right)^2=1\,{.}
\end{equation}
Eq. (\ref{isonew}) implies that
\begin{equation}\label{isocoord}
  r^*(r) = \int dr e^{(\nu-\mu)/2} = \int dr
  \left(1+\frac{B}{r}\right)^2\left(\frac{r-B}{r+B}\right)^{(C+1-\lambda)/\lambda}\,.
\end{equation}
Following the standard procedure (see for example \cite{Kof}), we
write the radial wave function in the form
 \[
 R(r^*)=\alpha(r^*)\beta (r^*)
 \]
and replace it into Eq. (\ref{isoR}). Then, one requires that the
coefficient of $d\alpha/dr^*$ vanishes, i.e.
\begin{equation}\label{isocoef}
\frac{d\beta}{dr^*}+G(r)\beta=0\,{,}
\end{equation}
where
\begin{equation}\label{isoeqG}
 G(r)\equiv \frac{e^{\nu/2-\mu/2}}{2} \left(\frac{2}{r}+\frac{\mu^\prime}{2}\right)
 \,{.}
\end{equation}
The integration of Eq. (\ref{isocoef}) yields the result
\begin{equation}\label{isobeta}
\beta (r^*)=\frac{\beta_0 e^{-\mu/4}}{r}\,,
\end{equation}
where $\beta_0$ is an integration constant. Taking into account
Eqs. (\ref{isonew}) and (\ref{isocoef}), Eq. (\ref{isoR}) leads to
the equation for $\alpha (r^*)$, whose form is Schroedinger--like, that is
\begin{equation}\label{sch}
-\frac{d^2\alpha}{dr^{*2}}+V_{eff}(r)\alpha=(\Omega^2-m^2)\alpha\,{,}
\end{equation}
where the effective potential for unitary mass $V_{eff}(r)$, i.e.
$V_{eff}(r)\to V_{eff}(r)/2m$, is given by
\begin{equation}\label{isoVeff}
V_{eff}(r)=\,G^2(r)+e^{\nu/2-\mu/2}\frac{dG(r)}{dr}
  + e^{\nu}\left[e^{-\mu}\,\frac{l(l+1)}{r^2}+m^2\right]-m^2\,{.}
 \end{equation}
To both sides of Eq. (\ref{sch}), we have add $-m^2$ in order that
the effective potential vanishes at infinity. $G(r)$ is defined in
Eq. (\ref{isoeqG}).

Near the singularity, i.e. in the spherical shell $r=B+x$, $x\ll
B$, the effective potential (\ref{isoVeff}) behaves as
\begin{eqnarray}\label{Veffx}
  V_{eff}(x)&\approx &
  \frac{1}{16}\left(\frac{x}{2B}\right)^{2(2+C)/\lambda}\frac{1}{x^2}
  \left\{
  \frac{B^2}{\lambda^2x^2}\, C\left(1-\frac{\omega C}{2}\right)
  +1+l(l+1)+{\cal O}(x)\right\} \\
  & \sim & 1.56\times 10^{-22}\frac{C}{\lambda}\left(1-\frac{\omega
  C}{2}\right)\left(\frac{{\rm Km}}{B}\right)^2
  \left(\frac{x}{2B}\right)^{2(2+C)/\lambda-4}\, {\rm eV}^2
  \nonumber
\end{eqnarray}
where Eq. (\ref{lambda}) has been used. Since $C<0$ or
$C>2/\omega$, as noted in the previous Section, the exponent of
the $x$-term is negative, so that the effective potential is an
infinite well for $x\to 0$. This behavior will be studied in detail in the next
Section.

It is worth to write down the effective potential obtained in the
framework of General Relativity, i.e. $C=0$ ($\lambda=1$),
 \begin{equation}\label{vgr}
  V_{eff}^{GR}=\frac{1}{(r+B)^8}\left\{
  B^5m^2(B^3+4B^2r+4Br^2-4r^3)+B^4r^2[l(l+1)-10m^2r^2] \right.
\end{equation}
 \[
  \left. -2B^2r^4[3+l(l+1)-2m^2r^2]+r^6[l(l+1)+m^2r^2]
  +2B^3r^3(1-2m^2r^2)+4Br^5(1+m^2r^2)\right\}-m^2\,,
 \]
which is finite at $r=B$.

Let us  now analyze the behavior of the effective potential
(\ref{isoVeff}) in the region $r>B$ and for different values of
parameters. Without loss of generality, we assume $B=5$ and $m=1$.

\begin{figure}[t]
\centering \leavevmode \epsfxsize=8cm \epsfysize=6cm\epsffile{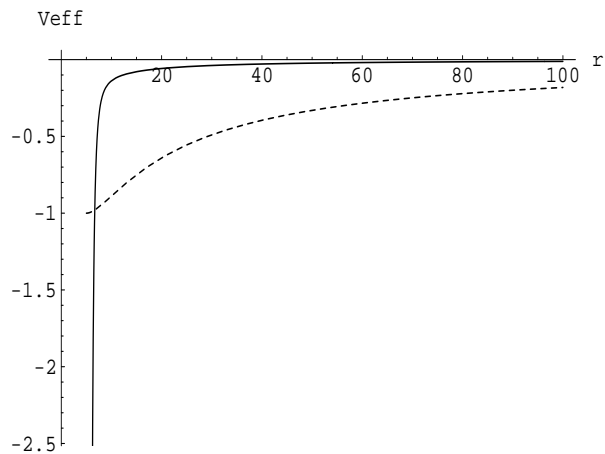}
\caption{Plot of the effective potential for $C=-1$, $l=0$, $\omega=600$. The solid line corresponds to
the effective potential (\ref{isoVeff}), the dashed line to (\ref{vgr}).} \label{BD+scalarFig1}
\end{figure}

\begin{figure}[t]
\centering \leavevmode \epsfxsize=8cm \epsfysize=6cm\epsffile{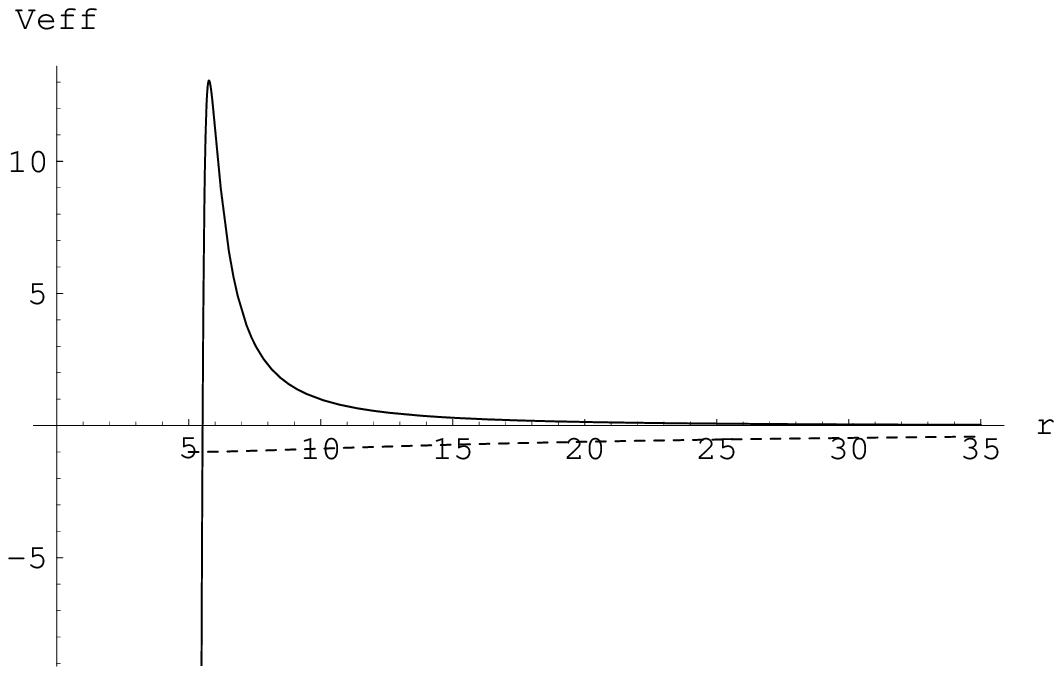}
\caption{Plot of the effective potential for $C=-1$, $l=8$, and $\omega=600$.
The solid line corresponds to the effective potential (\ref{isoVeff}), the dashed line to (\ref{vgr}).}
\label{BD+scalarFig2}
\end{figure}

\begin{figure}[t]
\centering \leavevmode \epsfxsize=8cm \epsfysize=6cm\epsffile{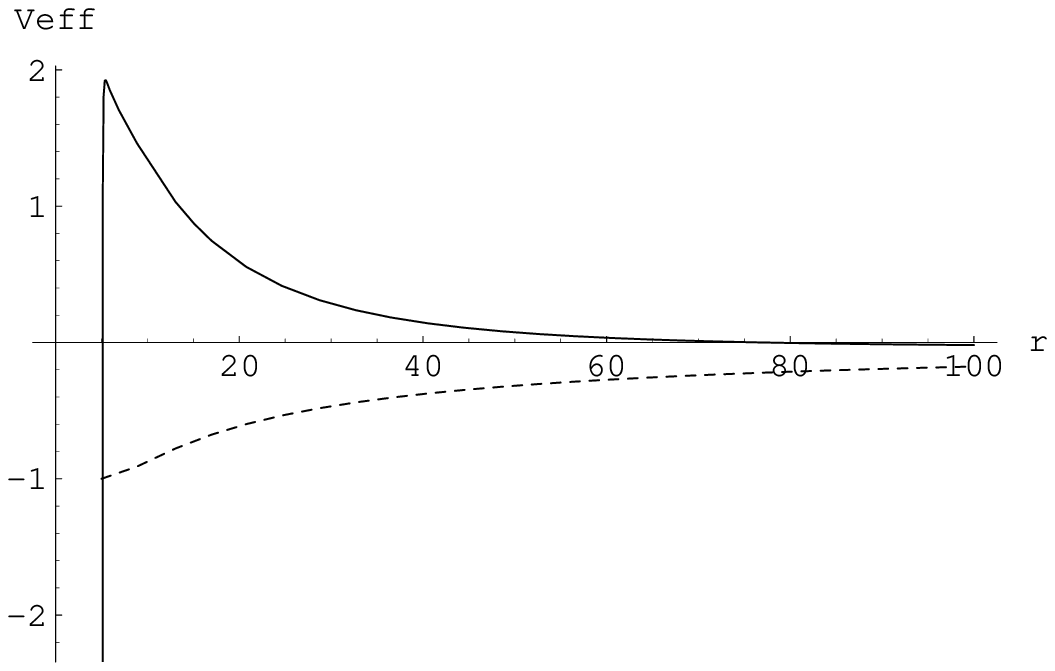}
\caption{Plot of the effective potential for $C=-0.1$, $l=30$, $\omega=600$.
The solid line corresponds to the effective potential (\ref{isoVeff}), the dashed line to (\ref{vgr}).}
\label{Veff+Ext+l=0+150}
\end{figure}

In Fig. \ref{BD+scalarFig1}, it  is plotted the effective potential (\ref{isoVeff}) for
radial motion, i.e. for vanishing angular momentum $l=0$.
The plot of the effective potential Eq. (\ref{isoVeff}) for non-vanishing angular momentum is reported in Fig. \ref{BD+scalarFig2}. Here $\omega=
600$, $C=-1$, and $l=8$ . The effective potential (\ref{isoVeff}), for large angular momentum and small $C$, is shown in Fig. \ref{Veff+Ext+l=0+150}.
In Figs. \ref{BD+scalarFig1}, \ref{BD+scalarFig2}, and \ref{Veff+Ext+l=0+150}, the quantities $r$ on the abscissa and
$V_{eff}$ on the ordinate are dimensionless ($r$/Km and $V_{eff}$Km$^2$, respectively). Besides, we also note that
$V_{eff}\to -\infty$ as $r\to 1$, whereas $V_{eff}^{GR}\to 0$. Obviously, for small values of the parameter $C$ ($C\approx 0$) the profiles of the effective potentials (\ref{isoVeff}) and
(\ref{vgr}) become equal.

It is worth stressing that we are only interested to values of the parameter $C$ belonging to the range $(-\infty, 0)\cup (2/\omega, +\infty)$,
since the scalar invariant are divergent, as it was shown in the previous Section.

\section{Radial Solution Near the Event Horizon}
\setcounter{equation}{0} Let us now investigate the behavior of
the radial part of the  wave function near $r=B$. This allows to simplify the
study of the radial differential equation given by Eq. (\ref{isoR}).
In the region near the surface $r=B$, it is possible to  write $r=B+x$, where
$x\ll B$. Replacing it into the radial Eq. (\ref{isoR}), one
obtains
\begin{equation}\label{eqx}
  y^{\prime\prime}+\left(1-\frac{C}{\lambda}\right)\frac{1}{x}\, y^\prime +
  \left[16(\Omega^2-m^2)\left(\frac{x}{2B}\right)^{2(\lambda-C-2)/\lambda}+
  m^2\left(\frac{x}{2B}\right)^{2(\lambda-C-1)/\lambda}-\frac{l(l+1)}{B^2}\right]
  y=0\,,
 \end{equation}
where $y\equiv R$, and the prime now indicates the derivative with
respect to $x$. In the limit $C=0$ and $\lambda=1$, the usual
result of the radial wave function in the isotropic coordinates is
recovered.

By introducing the variable $z=x/2B$, and taking $\Omega =m$,
$l=0$, Eq. (\ref{eqx}) assumes the form
\begin{equation}\label{eqz}
\frac{d^2y}{dz^2}+\left(1-\frac{C}{\lambda}\right)\frac{1}{x}\,\frac{dy}{dz}+
64B^2m^2 z^{2(\lambda-C-1)/\lambda}\, y=0\,,
\end{equation}
whose solution is \cite{riz}
\begin{equation}\label{soleqz}
  y = z^\alpha\left[C_1 J_\nu (\beta z^\gamma)+C_2Y_\nu (\beta
  z^\gamma)\right]\,,
\end{equation}
with
 \begin{equation}
 \alpha=\frac{C}{2\lambda}\,, \quad \gamma=2-\frac{C+1}{\lambda}\,,
 \quad \beta=\frac{8Bm}{\gamma}\,, \quad
 \nu=\frac{\alpha}{\gamma}\,.
 \end{equation}
In the limit $x\to 0$, the behavior of the Bessel functions is
\cite{riz}
 \begin{equation}
 J_\sigma (z)\sim \left(\frac{z}{2}\right)^\sigma \frac{1}{\Gamma(\sigma
 +1)}\,, \quad
 Y_\sigma(z)\sim -\frac{\Gamma(\sigma)}{\pi}\,
 \left(\frac{z}{2}\right)^{-\sigma}\,,
 \end{equation}
where $\Gamma (z)$ is the gamma function. Hence, the solution
(\ref{soleqz}) becomes
\begin{equation}\label{soleqz0}
  y\sim C_1+C_2 z^{2\alpha}\,,
\end{equation}
thus $P=|y|^2\to 1$ as $x\to 0$ (provided the wave function is
normalized, i.e. $C_1=1$). As expected, the above analysis
confirms results obtained in the previous Section, i.e. the
probability to find a scalar particle at the singularity $r=B$ is
finite since the effective potential has an infinite well.

\section{The case of  Gamma ray Bursts}
The previous results suggest a possible novel mechanism for the emission of
the intense beams of gamma rays. Accidentally discovered thirty
years ago, GRBs are the most energetic events in the Universe. They
are characterized by the following properties: $i)$  the bursts  are
isotropically distributed and are originated at cosmological
distance; $ii)$ they have an extraordinary large energy outputs ranging from  $\sim 10^{51}$erg/s to $10^{54}$erg/s; $iii)$  the spectra
are non-thermal, and, as widely believed, can be related  to synchrotron
radiation; $iv)$ their duration range from $10^{-3}$s to $10^3$s. Several 
models describe these phenomena, but the nature of GRBs
progenitor still remains obscure. Some models propose as
progenitors the merge of two neutron stars, of a neutron star and
a black hole or other compact objects: in general, all of them lead to a self-gravitating system
with a central black hole surrounded by a torus of dense matter.

Here, we are going to discuss  the emission of radiation generated by
charged scalar particles propagating in a BD gravitational field. We will show that, by such a mechanism, it is possible to achieve, in principle, the energies observed for GRBs. As we will see, such a feature depends on the BD parameters.

We
shall neglect some physical effects as the quantum particle
(pairs) creation induced by high space-time curvature or the
interaction of scalar particles with magnetic fields. Besides,
also the back-reaction effects will be neglected in our analysis,
being just interested to the emission of radiation induced by the
acceleration. It is expected that the two effects, i.e. the emission of
radiation by charged particles and the total energy of the
gravitational field generated by the BD scalar field though the
Einstein equation,  are comparable. In other words,  our propose is to suggest a
further mechanism which could  be taken into account in the context of 
GRBs physics.

The possibility that charged particles
may suffer high acceleration in the neighborhood of the singularity B would generate enormous
streams of photons by {\it brehmsstrahlung} (such a possibility has been already argued in Ref. \cite{21} for the
merging of neutron stars in a naked singularity). The power radiated away by a particle of
charge $e$ is estimated to be (in natural units) \cite{landau}
 \begin{equation}
W = -\frac{2e^2}{3} {\bf a}^2 \,,
 \end{equation}
where $|{\bf a}|$ is the modulus of the acceleration. In what follows, we shall consider charged
particles (e.g. fermions) described by scalar fields since spin effects play no role in the problem
at hand. As we said, other physical effects, as the interaction of particles with magnetic fields around
the gravitational source, the quantum particle (pairs) creation in high space-time curvature,
the back-reaction effects, and the total energy of the gravitational field generated by the BD
scalar field through the Einstein equation will be neglected, since we are interested, here,  only in
the emission of radiation induced by gravitational acceleration. 

The acceleration of charged particles is determined by the equation
 \begin{equation}\label{accGRB}
 {\bf a} = \frac{1}{i}[H, {\bf p}] = -\frac{1}{m}\left(\frac{dV_{eff}}{d r}\right) {\bf e}_r\,,
 \end{equation}
where $H$ is the Hamiltonian of the particle propagating the curved background,
$V_{eff} (r)$ is defined by Eq. (\ref{isoVeff}) and ${\bf e}_r$ is a versor directed along the
radial direction. Recalling that $V_{eff}$ is the effective potential for unitary mass, the acceleration
of the scalar charged particle near the singularity $r = B$ ($x \approx 0$) turns out to be
\begin{eqnarray}\label{amodulus}
   |{\bf a}|  \approx  \frac{B^2}{8 m_0^2} {\cal F}\left(\frac{x}{2B}\right)^{2(2+C)/\lambda}\frac{1}{x^5} 
   \approx  1.2\times 10^{-48}{\cal F} \left(\frac{\text{GeV}}{m_0}\frac{B}{\text{km}}\right)^2 \left(\frac{x}{\text{km}}\frac{\text{km}}{2B}\right)^{2(2+C)/\lambda}
   \left(\frac{\text{km}}{x}\right)^5 \text{eV}\,,
\end{eqnarray}
where
 \[
  {\cal F}\equiv \frac{C}{\lambda^3} \left(1-\frac{\omega C}{2}\right) \left(\lambda-1-\frac{C}{2}\right)\,.
 \]
The emitted energy by the charged particle is therefore
\begin{equation}\label{Wemitted}
    W\approx 10^{-97} {\cal F}^2 \left(\frac{\text{GeV}}{m_0}\frac{B}{\text{km}}\right)^4 \left(\frac{x}{\text{km}}\frac{\text{km}}{2B}\right)^{4(2+C)/\lambda}
   \left(\frac{\text{km}}{x}\right)^{10} \text{eV}^2\,,
\end{equation}
depending on the parameters of the BD spherically symmetric solution.
The mechanism for generating photons with high energy is strictly related to the (infinite)
potential well near the singularity $r = B$. The latter occurs owing to the presence of the
scalar field as source of the gravitational field. The issue is: Do naked singularities exist or
form during the gravitational collapse? According to the cosmic censorship hypothesis, the
formation of naked singularities is forbidden \cite{23}. Nevertheless, two comments are in order at this point:
i) the validity of this conjecture is still an open issue in General Relativity (there exist many
counterexamples that violate such a conjecture, see, for example, \cite{12,24}); ii) it is expected
that a full quantum theory of gravity may prevent the occurrence of singularity. At the
Planck scale, characterized by the scale length $l_P \sim 10^{-33}$cm, the quantum (gravitational)
effects become important, since they allow to solve the short-distance problems of General
Relativity by smearing out the classical singularities. Thus, at this distance, one has a finite
potential well, and the emitted energy is the maximum one.

By requiring that the maximum emitted energy is comparable with the observed GRBs
energy, i.e. $W \simeq  W_{GRBs}$, one can fix the parameter $C$ (actually, we are determining an upper
bound on $|C|$, since $W \lesssim W_{GRBs}$). Here $W_{GRBs} \sim  10^{54}$ erg/s $\simeq 10^{51}$ eV$^2$.
Estimations are carried out for protons, $m_0 \sim 1$ GeV, and $B \approx 3$ km (the mass source is of the order of the
solar mass). Taking into account all these features  and fixing $\omega = 5000$, we find that the parameter
$C$ turns out to be of the order $C \simeq  0.01815$. In the case where $\omega \sim 500$, the parameter
$C$ assumes the value $C \sim 0.055$. Therefore a small deviation from General Relativity (where $C = 0$) allows the production of the huge amount of energy, typical of GRBs.

For the sake of completeness, let us  evaluate the energy emission of charged particles near
the event horizon, i.e. in the framework of General Relativity. The acceleration is
$|{\bf a}|_{GR} \simeq \displaystyle{\frac{1}{128 m^2 B^3}\sim 10^{-51} \left(\frac{\text{GeV}}{m}\right)^2 \left(\frac{\text{km}}{B}\right)^3} \text{eV}$,
so that the emitted energy turns out to be
$W_{GR}\sim 10^{-102} \displaystyle{\left(\frac{\text{GeV}}{m}\right)^4 \left(\frac{\text{km}}{B}\right)^6}$,
 which is negligible in comparison to the observed energies of GRBs.
Finally, a comment is in order. The isotropy and homogeneity of matter distribution
in the Universe suggest that gravitational collapse or the emission of radiation by highly
accelerated charged particles generate bursts of photons that are isotropically distributed.
Thus, alternative theories of gravity seem to be a suitable framework for future investigations
of GRBs physics.

\section{Conclusions}

In this paper we have investigated the propagation of quantum particles in the framework of a spherically symmetric solutions achieved in the framework of 
BD theory of gravity. Such a propagation could explain  some features of GRBs physics. 

Specifically, we have shown that in the BD geometry, (scalar) particles near the
naked singularity suffer high accelerations, producing photons with an extraordinarily large
energy output (emission from naked singularity, the latter being regularized by quantum fluctuations
occurring at the Planck scale). This is in agreement with the fact that GRBs spectrum
is nonthermal, and, as widely believed, it is probably due to the synchrotron radiation. 

Conversely, GRBs could provide a direct method for testing strong gravity, which,
in turn, might provide a formidable tool in order to distinguish alternative theories of gravity from General
Relativity. Nevertheless,  a lot of  work is still necessary for a complete understanding of
the rough model here proposed. One of the basic issues is certainly represented by a careful analysis of the
back-reaction effects, including the radiation by BD scalar field, which could play a nontrivial
role on the structure of the mathematical formalism  used here.

Finally, we want to stress again that the BD theory is only a prototype of large families of models including scalar-tensor gravity with self-interaction potentials, $f(R)$ gravity, Gauss-Bonnet gravity \cite{felix}, hybrid gravity \cite{lobo} and so on describing a phenomenology which could give an alternative picture to dark energy and dark matter. Finding out new test-beds as GRBs  \cite{spyros} or anomalous compact objects \cite{neutron}, independent of the standard dark side phenomenology, could represent an extremely  interesting research line for these approaches.

\acknowledgments S.C. is supported by INFN ({\it iniziativa specifica} TEONGRAV). G.L. wishes to thank the Agenzia Spaziale Italiana
(ASI) for partial support through the contract ASI number I/034/12/0.

\end{document}